%% file: main.tex
\documentclass[conference]{IEEEtran}

\usepackage{xcolor}
\usepackage{graphicx}
\usepackage[compress]{cite}
\usepackage{hyperref}
\usepackage{ulem}
\usepackage{amssymb}%
\usepackage{pifont}%

\ifCLASSINFOpdf
\else
\fi

\usepackage{array}

\usepackage{url}

\newcommand{\pname}{Erudite}

\begin{document}
\title{Tearing Down the Memory Wall\vspace{-2ex}}

\author{\IEEEauthorblockN{Zaid Qureshi\IEEEauthorrefmark{2}\IEEEauthorrefmark{1}, Vikram Sharma Mailthody\IEEEauthorrefmark{3}\IEEEauthorrefmark{1}, Seung Won Min\IEEEauthorrefmark{3}\IEEEauthorrefmark{1}, I-Hsin Chung\IEEEauthorrefmark{4}, Jinjun Xiong\IEEEauthorrefmark{4}, Wen-mei Hwu\IEEEauthorrefmark{3}}

\IEEEauthorblockA{\IEEEauthorrefmark{1}Contributed Equally}

\IEEEauthorblockA{\IEEEauthorrefmark{2}CS, \IEEEauthorrefmark{3}ECE, 
University of Illinois at Urbana-Champaign, Urbana, IL 61801}

\IEEEauthorblockA{\IEEEauthorrefmark{4}Cognitive Computing Systems Research, IBM Thomas J. Watson Research Center, Yorktown Heights, NY, 10598}

\IEEEauthorblockA{
\{zaidq2, vsm2, min16\}@illinois.edu, \{ihchung,jinjun\}@us.ibm.com, w-hwu@illinois.edu}}

\maketitle

\input{abstract}

\input{intro}

\input{background}
\input{design}

\input{challenges}

\input{conclusion}

\input{acknowledgment}

\bibliographystyle{IEEEtranS}
\bibliography{ref}

\end{document}

%% file: abstract.tex
\begin{abstract}
We present a vision for the Erudite architecture that redefines the compute and memory abstractions such that memory bandwidth and capacity become first-class citizens along with compute throughput. In this architecture, we envision coupling a high-density, massively parallel memory technology like Flash with programmable near-data accelerators, like the streaming multiprocessors in modern GPUs. Each accelerator has a local pool of storage-class memory that it can access at high throughput by initiating very large numbers of overlapping request that help to tolerate long access latency. The accelerators can also communicate with each other and remote memory through a high-throughput low-latency interconnect. As a result, systems based on the Erudite architecture scale compute and memory bandwidth at the same rate, tearing down the notorious memory wall that has plagued computer architecture for generations. In this paper, we present the motivation, rationale, design, benefit, and research challenges for Erudite.
\end{abstract}

%% file: intro.tex
\section{Introduction}
\label{sec:intro}
The memory\footnote{In this paper, we use the term "memory" to broadly refer to both DRAM and storage devices that store and supply data for applications.} demands of emerging workloads such as Artificial Intelligence(AI)/Machine learning (ML), recommendation systems, high-resolution imaging, graph and data analytics is rapidly increasing~\cite{bert,gpt2, gpt3, dlrm,friendster,MOLIERE_2016,gpipe,resnet50,graphchallenge18,graphchallengetc19,graphchallengektruss19}.
For instance, the number of parameters needed to perform language modelling~\cite{bert,gpt3}, a common task in natural language processing, has grown by 507$\times$ over the last three years.
Furthermore, as many of these applications exhibit modest level of reuse over their large datasets, data movement becomes their primary performance bottleneck.
As a result, these emerging workloads require memory systems with large capacities and high bandwidth.

On the other hand, we are facing the ever widening gap between the raw compute throughput of modern computing systems and the memory capacity and bandwidth available to them.
Although there has been extensive development in compute engines, like GPUs~\cite{amperewhite} and TPUs~\cite{tpu}, memory technology has failed to keep up.
For example, the raw compute throughput of NVIDIA GPUs has increased by 30$\times$ over the last four years~\cite{pascalwhite,amperewhite}, while their memory capacity and bandwidth have only improved by 2.5$\times$ and 2.1$\times$, respectively.
In light of these trends, it is becoming increasingly difficult to make effective use of the compute capability of modern computing systems for emerging workloads.

To alleviate the impact of limited memory capacity in compute devices like GPUs, modern computing systems deploy fast NVMe SSDs and rely on the application and the operating system (OS) running on the CPU to orchestrate the data movement between the GPUs, CPU memory and SSDs, while supporting standard abstractions such as memory-mapped files~\cite{flatflash,dragon}.
Using such a design to manage massive datasets comes with performance challenges that arise from the infrastructure software overhead, long latency of SSD access, and limited number of parallel SSD requests that the CPUs can initiate in modern computing systems. 

The traditional solution is to use coarse-granular data transfers so that the software overhead, SSD access latency, and limited request generation throughput can be amortized with a large number of bytes transferred for each request. As a result, the transfer size for each request has been ranging from 4K bytes to multiple mega bytes in modern computing systems. In fact, many applications even transfer the entire data set to the compute engine memory and let the computation pick out the portions of the data to be used.  The assumption is that the large amount of data thus transferred are well utilized by the compute engines. That is, the data access patterns exhibit a high-level of spatial locality. 

Unfortunately, such assumption has become less and less valid for modern analytics and AI applications where data lookups are increasingly data-dependent and sparse in nature~\cite{rapids,dlrm,emogi}. Intuitively, the key to efficiently analyzing a massive data set is to strategically touch as little of the data as possible for each application-level query. As a result, coarse-grained accesses have increasingly resulted in unused data being transferred, a phenomenon commonly referred to as I/O amplification. 

I/O amplification can significantly reduce the effective memory bandwidth for data actually used by the compute engine. Since the available bandwidth for SSD accesses is already orders of magnitude lower than DRAM/HBM, such reduced effective bandwidth can severely impact the overall performance of applications.  It is thus desirable to reduce the data transfer granularity to preserve as much of the effective access bandwidth. For example, using 128-byte to 512-byte data transfer granularity can dramatically reduce I/O amplification in graph traversal workloads~\cite{emogi}.

However, with finer-grained data transfers, one can no longer amortize the software overhead, SSD access latency, and limited request generation throughput over a large number of bytes being transferred. For example, assume a system with 16GBps PCIe one-way bandwidth, 64$\mu$s SSD access latency, and 512-byte data transfer granularity. Each data transfer will only occupy the PCIe link for $512$B$/16$GBps $= 32$ ns. In order to tolerate the 64$\mu$s latency and sustain the theoretical 16GB/s data access throughput, the system must be able to initiate an access every 32ns and maintain 64$\mu$s/32ns = 2,000 simultaneous accesses at any given time. Obviously, the required rate of requests will increase as the granularity decreases and the bandwidth of the interconnect increases. Such a high rate of request cannot be achieved with current combination of limited parallelism in the CPU hardware and the traditional file system software.

Thus, in a system where all SSD accesses must be initiated by the CPU, the CPU will most certainly become the bottleneck when it comes to requesting fine-grained data transfers on-behalf of the fast compute engines like GPUs. %
This not only limits the performance of the GPU but also the NVMe SSDs as shown by prior works~\cite{flatflash,flashshare,dragon}. 
Thus, to improve performance, we propose to move the slow CPU and its software stack out of the GPU's access path. 
The idea is to allow the massive number of GPU threads to directly initiate fine-grained NVMe accesses, thus matching the SSD data transfer parallelism to the compute parallelism of the GPUs. %
A more subtle rationale for enabling GPU threads to directly initiate NVMe SSD accesses is to allow the data access decisions to be seamlessly made by the algorithm code being executed in the GPU. Such integration allows the applications to make fine-grained, data dependent accesses requests in a natural way, thus better meet the needs of modern data analytics and AI applications.

This paper presents the \textit{\pname{}} system architecture that enables compute engines to perform massively parallel, fine-grained accesses to the SSD to tackle the memory wall and scalability challenges in modern computing systems. 
\pname{} comprises of multiple \pname{} Processing Units (EPUs) interconnected by the low-latency and high-bandwidth \pname{} switch. 
In \pname{}, an EPU is the mechanism to scale the compute and memory capacity/bandwidth at the same rate.
Each EPU consists of a compute unit based on the GPU architecture called the \pname{} Compute Unit (ECU), a few gigabytes of high-bandwidth memory (HBM), and an array of storage-class memory like NVMe SSDs.
In contrast to the conventional CPU-centric system design, in \pname{} the ECU is the root for a tree of SSDs, providing aggregate performance of the SSDs to the compute unit. 
This enables the \pname{} design to match the compute throughput to the data access bandwidth and also brings the compute near the data. 

We enable the ECU to have direct and secure access to the NVMe SSDs by designing the \pname{} controller and integrating it into the EPU.
The \pname{} controller virtualizes the SSDs of an EPU and provides its own NVMe interface for the ECU.
Initially, the CPU maps this NVMe interface into ECU's  memory and the application's address space, enabling the compute threads to directly make requests to the \pname{} controller for data on the backing SSDs.
The \pname{} controller runs a lightweight file system for organizing data on the SSDs and provides the needed access control for accesses to the NVMe SSDs. 
On each request, the file system checks if the requesting application has access to the requested data.
Thus, \pname{} provides secure access to data and removes the CPU and OS from both data and control planes entirely.

We are currently in the process of building \pname{}.
However, there are many system challenges that we need to overcome as many components like systems software and interconnects are designed for CPU-centric architectures. 
We discuss some of these challenges in detail and propose potential solutions for them. 
We hope to foster discussion within the computer systems community about developing a scalable architecture where both compute and memory are scaled equally. 

The rest of the paper is organized as follows. In $\S$~\ref{sec:relatedwork} we take a deep dive into the current application trends, discuss limitations of conventional CPU-centric architectures and motivate why we need to re-design the system architecture.
We present the \pname{} architecture and its components in $\S$~\ref{sec:design}.
The challenges we face in building the \pname{} architecture are detailed 
in $\S$~\ref{sec:challenge}.
We conclude in $\S$~\ref{sec:conclusion}.

%% file: background.tex
\section{Background and Motivation}
\label{sec:relatedwork}
In this section, we discuss emerging application trends, the challenges in modern computing systems and provide a brief overview why Flash should be used as memory.

\subsection{Emerging Application Trends}
\begin{figure}[t]
  \centering
  \includegraphics[width=\linewidth]{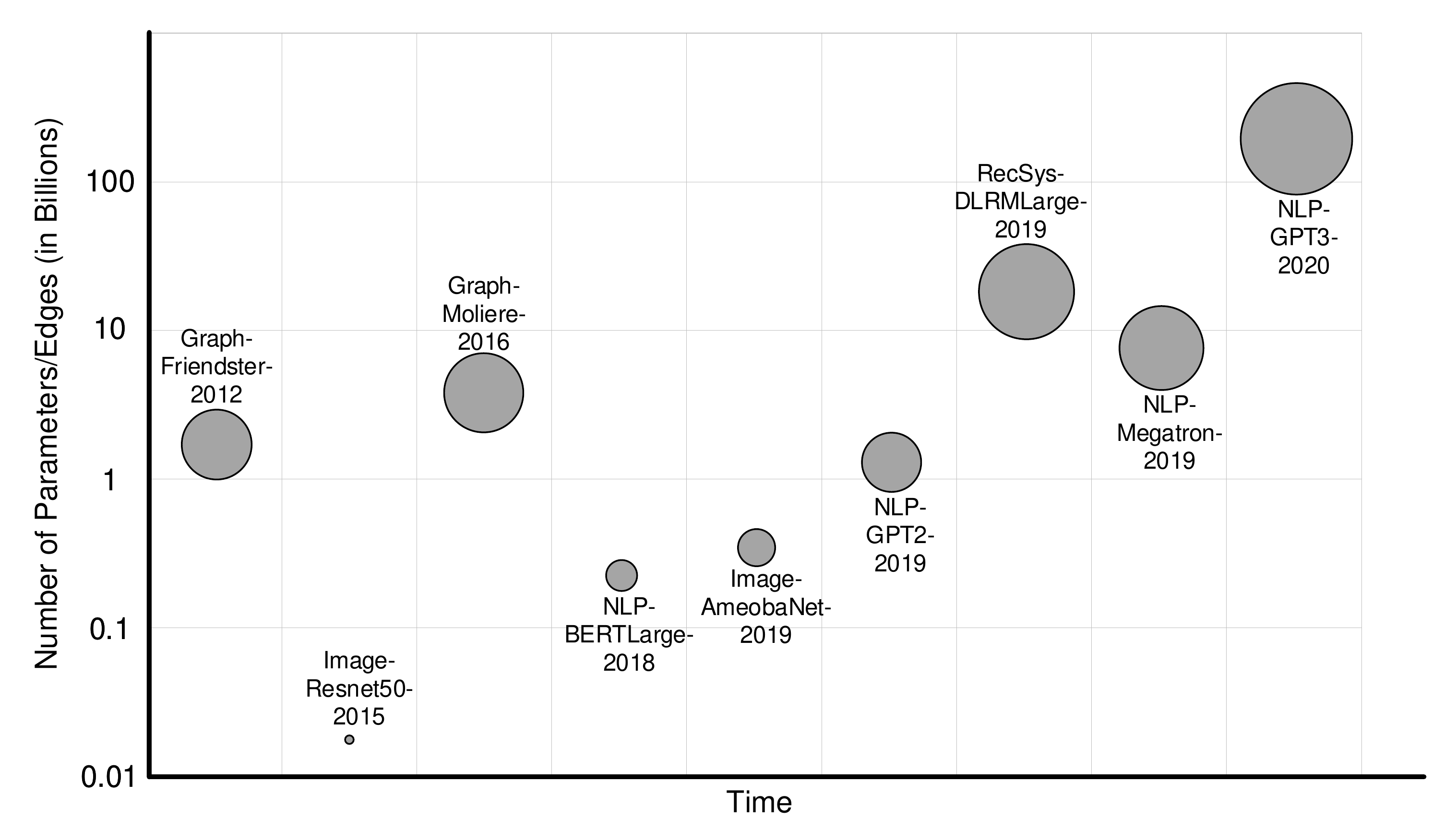}
  \caption{Trend in the number of parameters for applications in three different domains. Taking the NLP application~\cite{bert,gpt3} as an example, the number of parameters have increased by 507$\times$ in just three years. The number of parameters is directly proportional to the application's memory requirement.}
  \label{fig:capacity}
\end{figure}

Many emerging applications such as graph and data analytics, recommendation systems, natural language processing and high-resolution imaging require very large amount of memory for their efficient execution. 
Figure~\ref{fig:capacity} shows the trends in the number of parameters used in deep learning models and the number of edges in graphs over 8 years for image recognition~\cite{resnet50,gpipe}, natural language processing~\cite{bert, gpt2, megatronlm, gpt3}, recommendation systems~\cite{dlrm} and graph analytics~\cite{friendster,MOLIERE_2016} applications. 
Consider natural language processing, the number of parameters has grown from 345 Million in the BERT-Large model~\cite{bert} to 175 Billion parameters in GPT3 model~\cite{gpt3} over the span of just three years. 
The number of parameters and the number of edges are directly proportional to the memory footprint of deep-learning and graph analytics applications, respectively.
This trend shows that the application memory footprint is expanding at an unprecedented rate.%

Some emerging applications such as recommendation systems~\cite{dlrm} and GPU-accelerated data analytics~\cite{rapids, emogi} over datafames exhibit irregular memory access patterns which further constraints the memory system. 
Common operations such as join, filtering or embedding table look-up require data-dependent accesses. 
Given that the dataframes in data analytics applications have millions of rows, these accesses can become sparse during such operations.
Sparse irregular accesses can severely degrade the performance of the system as the current hardware is designed to exploit spatial locality.

\subsection{Conventional System Architecture}
\label{sec:conv_hard_design}
\begin{figure}[t]
  \centering
  \includegraphics[width=0.8\linewidth]{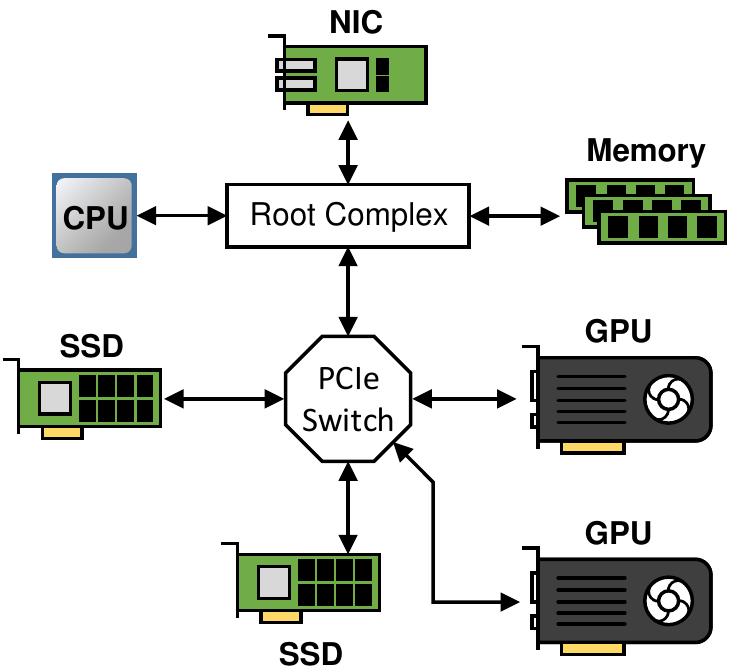}
  \caption{Conventional system architecture. Here the CPU is the master of all compute, memory, and storage devices and orchestrates the data movement between them.}
  \label{fig:baseline_design}
\end{figure}

\textbf{Hardware:} Figure~\ref{fig:baseline_design} shows the conventional system architecture widely used in today's data centers and super computers. 
Multiple accelerators and storage devices are connected to the CPU root complex trough a PCIe switch.
Such a design was convenient in the past as it allowed the CPU to be the master of all devices and  orchestrate data movement. 
However, over the last decade, accelerators such as GPUs have become the primary compute engines. %
Using the CPU-centric architecture in this new paradigm fundamentally limits the performance of these heterogeneous systems. 
This is because applications executing on GPUs do not have control over the storage data access and require constant interactions with the CPU to fetch the needed data. %
As discussed in $\S$~\ref{sec:intro}, the CPU hardware and the traditional file system software cannot initiate fine-grained data transfer requests at sufficiently high rate to take full advantage of the current and coming generations of interconnects and SSDs and meet the data consumption needs of the GPUs. 

\begin{figure}[t]
  \centering
  \includegraphics[width=0.9\linewidth]{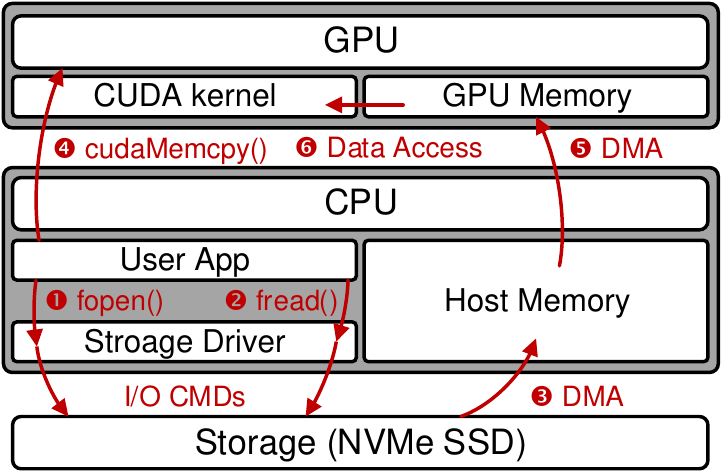}
  \caption{Conventional storage access flow when using a GPU. In the traditional system, the CPU manages both the storage and the GPU's memory and thus only the CPU can securely move data between the storage and the GPU.}
  \label{fig:access_flow_baseline}
\end{figure}

\textbf{Software:} In addition to the poor throughput and scalability offered by the conventional hardware design, the software overheads from page faults, system calls and file system are huge.
Let us consider an application which requires the GPU to process data stored in the SSD.
Figure~\ref{fig:access_flow_baseline} describes a read operation from user application that occurs in the heterogeneous system consisting of NVIDIA GPU for compute and NVMe SSD for storing large datasets.
When the user application opens a file with read permission (\ding{182}), the operating system checks if the user has correct permissions to access the file. 
If it does, the userspace application is allowed to make read and write requests for the the data on the NVMe SSD. 
These read and write operations (\ding{183}) require the operating system to 1) check access permissions to specified memory region, 2) traverse multiple layers of indirection to find the right logical block address (LBA) for the data, and 3) create a NVMe I/O command packet that is then sent to the SSD using NVMe driver. 

NVMe SSD reads the data from the Flash device based on the information provided in the NVMe I/O command and copies (\ding{184}) the data to the pre-allocated and pinned host memory using Direct-Memory Access (DMA).
After the data is copied to a buffer in the application's address space, the application can copy the data from the CPU memory to the GPU memory (\ding{185}) using DMA (\ding{186}).\footnote{Steps 3, 4, and 5 can be combined into a copy from the SSD into the GPU memory using GPUDirect storage access capability~\cite{gds}.}
Now the GPU's CUDA kernel is allowed to access the data for its computation (\ding{187}). 
The entire timeline of execution in the conventional CPU-centric design is summarized in Figure~\ref{fig:bar_flow}(a).
Previous works have shown that this stack has significant performance overheads~\cite{ flatflash,flashshare}.

The above example assumes that GPU threads only work on the data in the GPU memory and don't make any accesses to data not in the GPU memory.
Of course large working sets cannot fit in the physical memory of modern GPUs, so recent work has enabled GPUs to access data in CPU memory~\cite{uvm} and even in memory-mapped files~\cite{dragon} through the page-faulting mechanism.
Such a mechanism is crucial to enable GPU applications like graph and data analytics to make data-dependent accesses over large data-sets.
With such a system, when a GPU thread tries to access some piece of data not in the GPU memory, a GPU page fault is triggered.
These page faults are handled by the GPU driver and operating system running on the CPU as they need to check permissions for access and perform the required data movement between the SSD and the GPU. 
If millions of GPU threads trigger page-faults, the CPU can easily get swamped and become the primary performance limiter for overall performance.

\label{sec:conv_soft_stack}

\subsection{Compute and Memory Gap}
\label{sec:computememgap}
\begin{figure}[t]
  \centering
  \includegraphics[width=\linewidth]{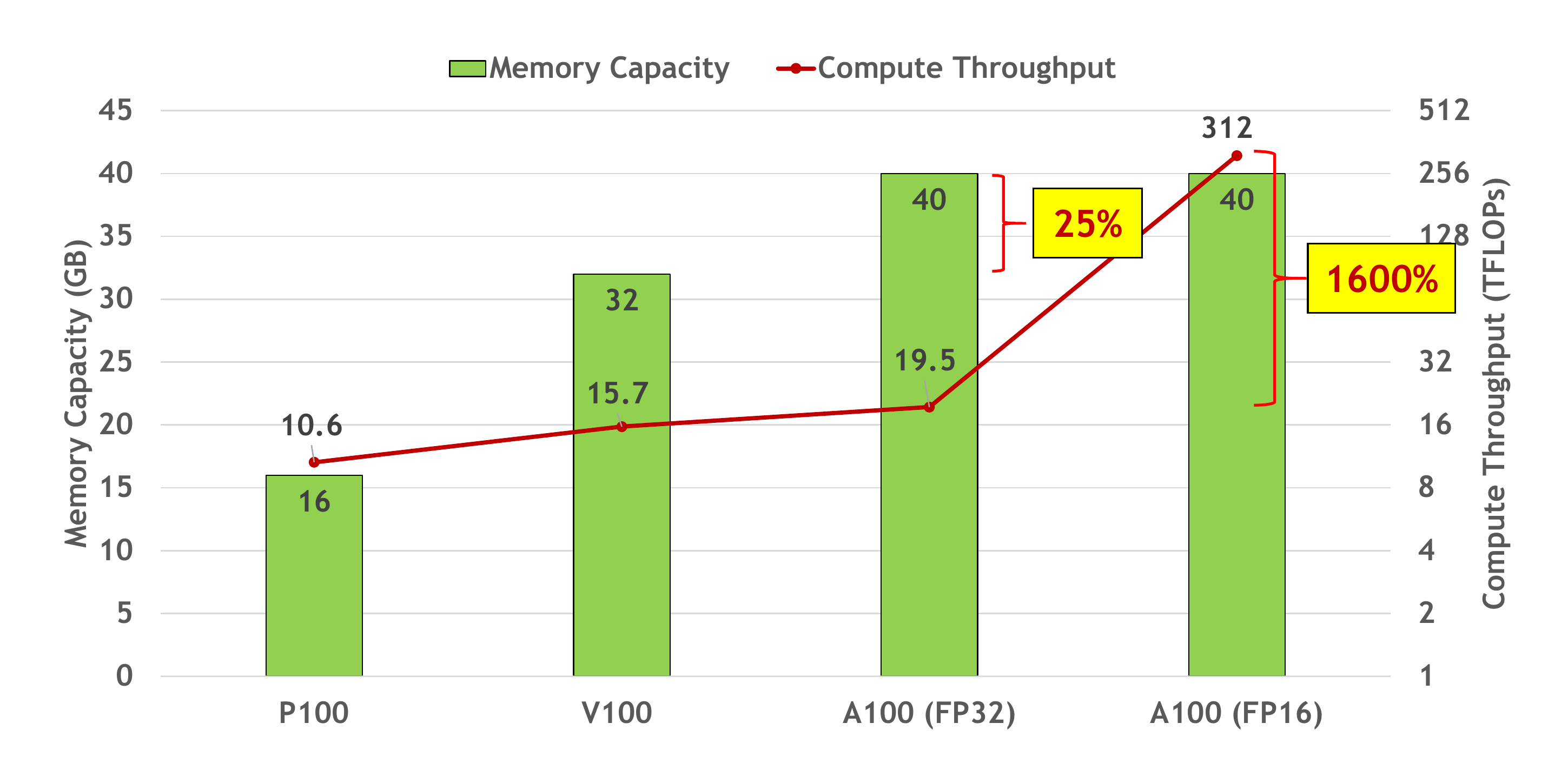}
  \vspace{-5ex}
  \caption{Compute and memory capacity trends in high-end NVIDIA GPUs over the last four years. Computational throughput has improved by almost 30$\times$ while memory capacity has only improved by 2.5$\times$.}
  \label{fig:memcompute}
\end{figure}

Even though clock frequency scaling has ended, architectural advancements like multi and many core architectures, SIMD and SIMT execution, computing on mixed precision, and throughput oriented computing have enabled the continued scaling of peak arithmetic throughput over the last decade~\cite{erudite}. 
Considering only the last four years of high-end NVIDIA GPUs, the Pascal P100~\cite{pascalwhite} provided peak single precision arithmetic throughput of 10.6TFLOPs while the Ampere A100~\cite{amperewhite} can achieve up to 312TFLOPs of compute operations with its half precision data format (FP16 - no sparsity), as shown in Figure~\ref{fig:memcompute}.
The compute throughput has improved by 30$\times$ in just four years.

Such dramatic increase in compute throughput has resulted in tremendous pressure on the memory subsystem since both memory capacity and bandwidth have failed to scale at the same rate. 
The P100 GPU~\cite{pascalwhite} had a memory capacity of 16GB and a memory bandwidth of 732GBps in 2016,
while the latest A100 GPU~\cite{amperewhite} has a memory capacity of 40GB and a memory bandwidth of 1555GB/s.
Memory capacity improved by only 2.5$\times$ between P100 and A100 GPUs and the memory bandwidth improved by only 2.1$\times$ in the last four years. 
Furthermore, if an application exhibits no-reuse, the P100's memory bandwidth only supports 186GFLOPs and the A100's memory bandwidth supports 389GFLOPs and 777 giga FP16 operations per second, an improvement of only 4.18$\times$ in four years.
In order for an application to achieve peak arithmetic throughput of the P100 and A100 GPUs, it would have to re-use each data item fetched from the GPU memory 56.9 times and 400 times, respectively.

If data does not fit in the GPU memory, then the re-use factor required to sustain peak arithmetic throughput significantly increases as accessing the external memory is constrained by the limited interconnect bandwidth and the slow CPU\cite{erudite}. 
Assuming the latest NVLink interconnect that can provide the A100 GPU 300GB/s of bandwidth to the CPU memory~\cite{amperewhite}, each FP16 data item fetched over this interconnect must be re-used 2080 times to sustain the peak compute throughput of the GPU. %
If the data does not fit in the host memory, then the access bandwidth between the GPU and SSD is limited to 32GB/s~\cite{pcisig}, requiring each fetched data item to be re-used 19,500 times to sustain the peak arithmetic throughput of the GPU.
Emerging applications don't exhibit this order of reuse, and thus the compute resources of GPUs are heavily underutilized.
Note that the overhead added by the CPU and its software stack can dramatically reduce the effective bandwidth and thus make the sustainable GPU compute throughput much worse.

\subsection{Flash as Memory}
\label{sec:flash_mem}
Because DRAM technology has limited density, industry and academia
have been exploring denser non-volatile storage class memories 
such as 3DXpoint~\cite{intel_optane_website}, MRAM, Flash and PCMs.
Storage class memories (SCM) are the prime candidate for next generation data centers 
main memory as they provide large memory capacity and offer low cost-per-bit %
compared to DRAM. %
SCM come in different types based on the technology used and they 
differ in size, density, latency, throughput, power and cost. 
An ideal SCM for tomorrow's data center has TBs of memory capacity, 
provides at worst microsecond latency, offers high memory bandwidth,
has very low power consumption, offers high memory parallelism, and is very cheap.
 
Unfortunately, none of the existing SCM in literature and products satisfy all the requirement 
of ideal SCM needed for tomorrow's data center main memory.
The new 3DXPoint memory technology provides an access latency of less than 1$\mu$s but is only about four times denser than DRAM at a density of 0.62Gbit/mm$^2$~\cite{optanedensity} compared to DRAM's density of 0.132Gbit/mm$^2$~\cite{dramdensity}.
Furthermore, it suffers from limited memory size, DRAM like energy consumption, and
high costs at \$6.75/GB~\cite{optaneprice}.
In fact the only storage class memory that satisfies most of the requirements is the mature Flash memory technology. 
Flash offers high throughput, thanks to several levels of parallelism, 
provides high density at 4.34Gbit/mm$^2$~\cite{optanedensity}, consumes very low power per memory access,
and is very cheap at \$0.1/GB~\cite{flashprice}. 

Using Flash based SSDs as part of main memory has been shown to be a practical 
approach to address the memory capacity requirements for data intensive applications~\cite{flatflash}. 
This is because Flash random access latency has now been reduced to just few tens of microseconds~\cite{znand,issccznand}.
State-of-the-art systems support large memory capacities by memory-mapping SSDs and leveraging the
paging mechanism to access the data. 
However, such a scheme suffers from long latencies and low throughput due to high overhead from the software stack. 
Prior work such as~\cite{flatflash,scmdesign,2bssd} has tackled this high 
overhead by enabling cache line access to the SSD~\cite{2bssd, flatflash}, using host DRAM as a cache~\cite{scmdesign}, merging multiple translation layers 
into one layer, and promoting pages from SSD when locality is detected~\cite{flatflash}.
Due to its benefits, Flash memory is abundantly available, extensively used, and continues to be optimized to provide better performance~\cite{znand,issccznand,snapread}.

%% file: design.tex
\section{Design}
\label{sec:design}
\begin{figure}[t]
  \centering
  \includegraphics[width=\linewidth]{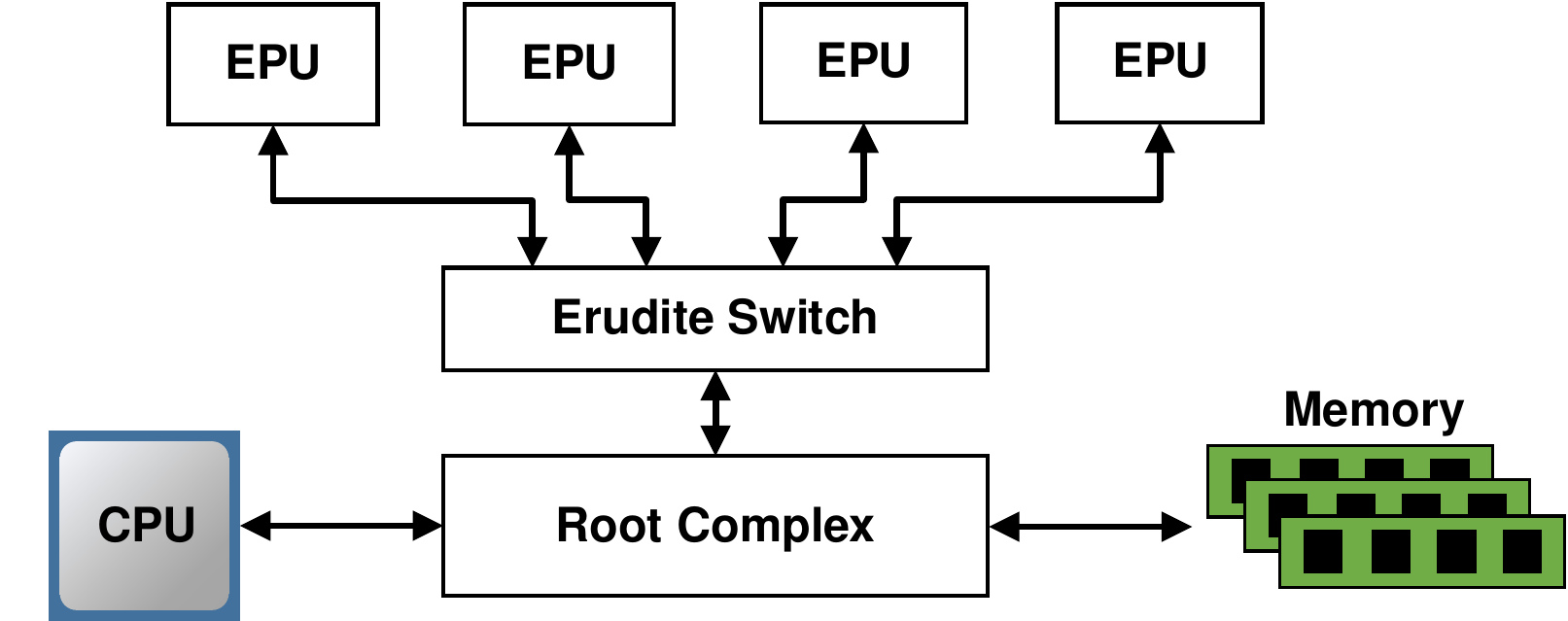}
  \caption{\pname{} system architecture. The \pname{} system consists of multiple \pname{} Processing Units (EPU) connected over the \pname{} Switch. This enables efficient scaling of not only compute but memory bandwidth and capacity as well.}
  \label{fig:design}
\end{figure}

We address the memory wall challenges of modern computing systems with \pname{}. An overview of the \pname{} system architecture is shown in Figure~\ref{fig:design}. At the core of the \pname{} system design is the \pname{} Processing Unit.
We scale both the memory capacity and compute capability in \pname{} by having multiple \pname{} Processing Units interconnected with a low-latency and high-bandwidth switch.
In this section we describe each of these components in depth and how they address the bottlenecks in conventional systems.

\subsection{\pname{} Processing Unit (EPU)}
\begin{figure}[t]
  \centering
  \includegraphics[width=\linewidth]{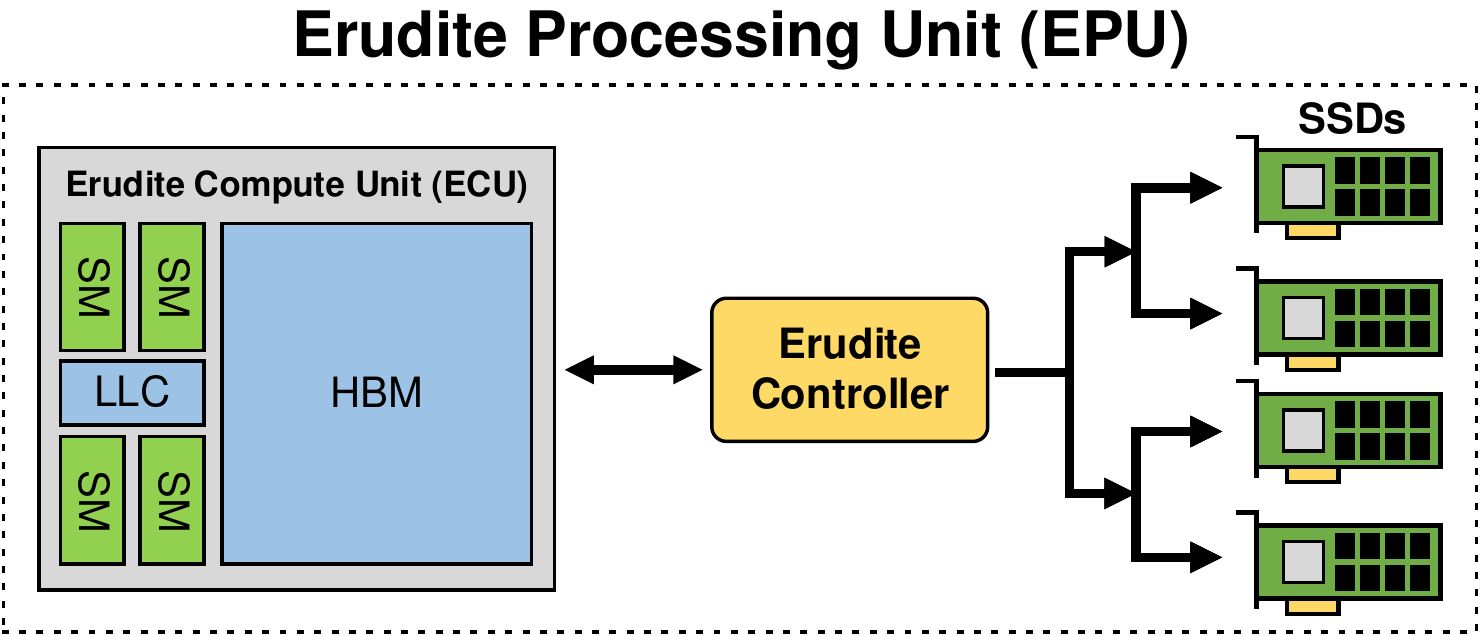}
  \caption{\pname{} processing unit (EPU) consists of a \pname{} Compute Unit, an array of NVMe SSDs, and a \pname{} Controller providing secure access to the NVMe SSDs.}
  \label{fig:epu}
\end{figure}

The EPU, shown in Figure ~\ref{fig:epu}, consists of a compute device, an array of SSDs as storage class memory, and an \pname{} controller providing secure access between the two.

\textbf{Compute: }
The EPU's compute device, the ECU, is based on the GPU architecture.
GPUs are designed to hide memory latencies by providing massive parallelism. In fact, recent work~\cite{emogi} has shown that with efficient scheduling, GPUs can hide latencies of accessing data across interconnects like PCIe and fully saturate the interconnect bandwidth even for irregular applications like graph traversal.
However, as noted in $\S$~\ref{sec:computememgap}, modern GPUs significantly mismatch the compute throughput they offer with the memory-bandwidth available to them. 
As such, unless one's application maps perfectly to something as compute-bound as matrix multiplication, it is very hard to reach the full performance of modern GPUs.

To circumvent this deficiency, \pname{} balances the compute capability of each \pname{} Processing Unit with the memory bandwidth available to it.
A GPU consist of multiple Streaming Multiprocessors (SM), each with up to 64 processing cores and the ability to schedule among 2048 concurrent threads at a time.
We reduce the number of parallel Streaming Multiprocessors (SM) in the ECU. 
We do this based on the insight that the ECU only needs the number of threads  that can fully utilize the memory bandwidth.

\textbf{Memory Hierarchy: }
Each EPU has a few gigabytes of HBM memory.
This HBM memory provides low-latency and high-bandwidth access to data structures with high levels of reuse and locality.
Each EPU also comes with an array of Flash SSDs as a cost and power efficient way to increase the EPU's memory capacity.
As noted in $\S$~\ref{sec:flash_mem}, Flash SSDs provide terabyte-scale memory capacities through Flash chip and channel parallelism.
A high-end Flash SSD can contain up to 32 Flash channels with up to 8 chips per channel~\cite{deepstore}.
This not only provides a large memory capacity, but a large amount of bandwidth as well.
High-end Flash SSDs can provide over 6GB/s of random access bandwidth over PCIe~\cite{samsungfast}.
To meet the bandwidth demand of the ECU, we scale up the number of Flash SSDs per ECU, allowing the ECU access to a large memory capacity and bandwidth.

Here we note a key difference between Erudite's design and the design of a conventional system. As noted in $\S$~\ref{sec:conv_hard_design}, the conventional system is CPU-centric and is designed in such a way that it is very difficult for an accelerator like a GPU to have access to the aggregate bandwidth of many PCIe-based SSDs. In Erudite, each ECU is the root for a tree of SSDs, providing the aggregate performance of the SSDs to the ECU. 

\textbf{\pname{} Controller and Software: }
High-end Flash SSDs provide non-volatile storage as well as high-performance over the NVMe protocol~\cite{nvme}.
In traditional systems, these SSDs are managed by drivers and file systems in the operating system running on the CPU, as described in $\S$~\ref{sec:conv_soft_stack}. 
As a result, in a traditional system there is a significant software stack that runs on the CPU to manage the GPU's memory, which can lead to large overheads.
In \pname{}, we remove the traditional, CPU-based operating system from both the data and control paths of NVMe SSDs, and bring NVMe control to the user-space application running on the EPU, akin to SPDK in the CPU user-space~\cite{spdk}. Moving NVMe control to the ECU-space allows the control to be performed in a massively parallel manner by threads running on the ECU.
We enable this by mapping NVMe queues and IO buffers in the ECU's HBM for peer-to-peer RDMA and mapping NVMe controller registers into the ECU's application address space.
This not only provides the ECU threads with low-latency, high-throughput access to NVMe SSDs, but also gives the programmer more control over the use the EPU's resources.
However, bringing NVMe control to user-space generally means there is no file-system managing the SSDs' non-volatile storage or providing access control.

In \pname{}, we aim to overcome this limitation and provide both of these features through the design and implementation of the \pname{} controller.
The \pname{} controller virtualizes the SSDs of an EPU, and provides its own NVMe interface for the ECU. 
The ECU's programming model is very similar to the CUDA programming model, allowing the expression of massive parallelism with many threads.
When the ECU threads want to issue reads or writes to the EPU controller, they can enqueue commands to the NVMe queues, ring the NVMe doorbell and poll for the command completion. 
The \pname{} controller will read these commands and check if the application issuing the requests has the permission to access the requested blocks of data. 
If the permission checks pass, the command is appropriately forwarded to the backing SSDs and data is transferred between the HBM and the SSDs directly.
When the command completes, the \pname{} controller will write the completion entry which the EPC threads are polling for.
Effectively, the \pname{} controller runs a lightweight file system, with optional hardware acceleration, for organizing the data on the SSDs and providing permission checks for accesses to the non-volatile SSDs.

\begin{figure}[t]
  \centering
  \includegraphics[width=\linewidth]{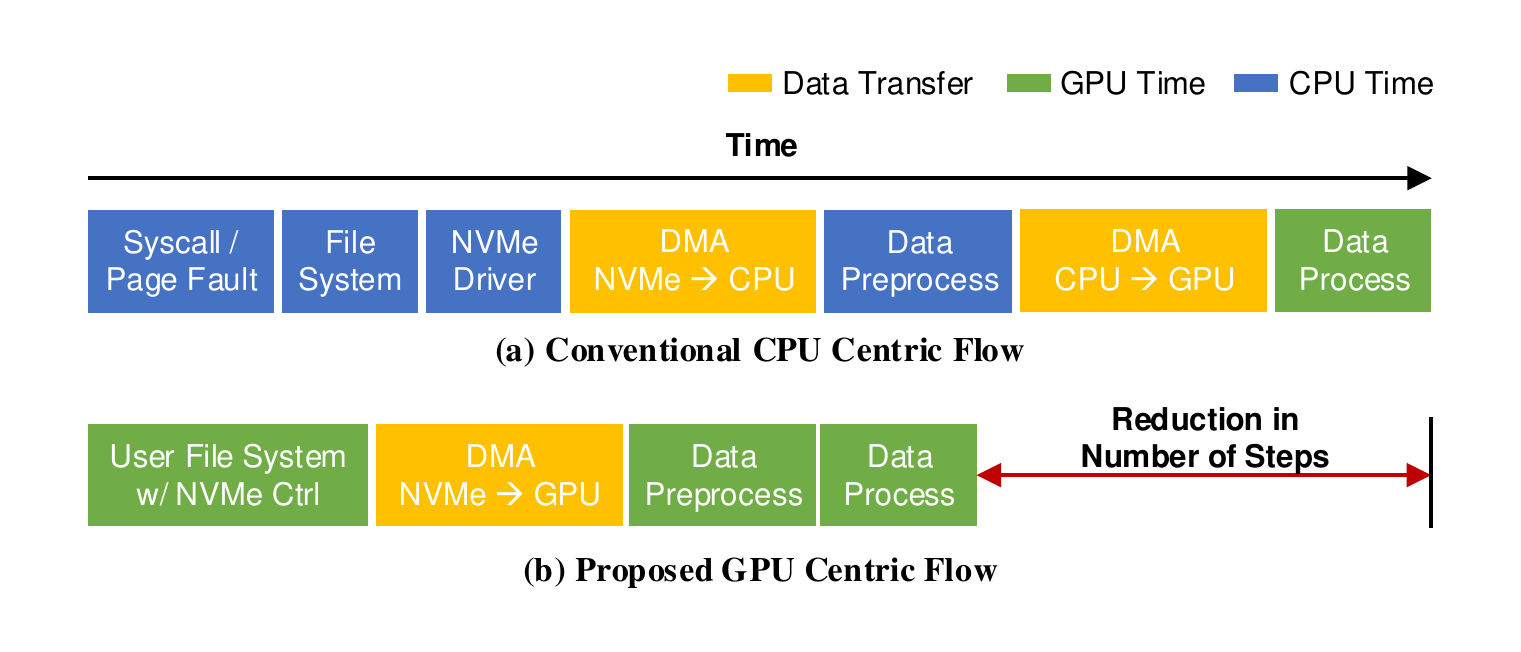}
  \caption{Steps involved for a GPU kernel while accessing data from the NVMe SSD in (a) conventional CPU-centric architecture and (b) \pname{} system architecture. Reducing the number of steps can significantly improve the number of parallel accesses the GPU can make to the NVMe SSD.
  }
  \label{fig:bar_flow}
\end{figure}

With such a design, \pname{} provides secure access to data without the involvement of the CPU or the operating system, as described in $\S$~\ref{sec:conv_soft_stack}.
We show the advantage of such an approach in Figure~\ref{fig:bar_flow}, where the CPU time is completely removed during application execution in \pname{}.
Furthermore, such a design allows the programmer to exploit more parallelism by enabling the programmer to have millions of NVMe requests in flight, which the protocol was designed to support~\cite{nvme}.
Finally, the programmer can manage their own memory, as in what is cached in the HBM, removing traditional performance issues like unwanted memory pollution and thrashing.

\subsection{\pname{} Switch}
To enable the scalability of both compute and memory, one of the major goals of \pname{}, we propose the \pname{} switch allowing multiple EPUs to be connected to the same system.
This switch enables two key features of \pname{}.
First, it enables the programmer to exploit data-level parallelism and near-data processing in their emerging bandwidth-limited applications.
We studied such an application, Intelligent Queries, in our previous work DeepStore~\cite{deepstore}, and found that near-data processing and data-level parallelism can enable the performance of these applications to scale linearly with the number of memory and compute elements.
In \pname{}, the switch enables multiple EPUs to exploit near-data processing and data-level parallelism with the ECUs and the local SSD arrays.
Moreover, the switch enables communication between multiple EPUs through a memory interface. 
This allows an EPU to access any other EPU's SSD array, as all the programmer needs to do is map the NVMe queues of one EPU \pname{} controller into another EPU's HBM.
The \pname{} controller will check the requests issued by applications on these queues just like it would for its local ECU, providing all compute units secure access to data in any EPU in the system. The Erudite Switch differ from existing industry switch architectures in two aspects. First, the header is designed for fine-grained transfers. Second, the tags are much larger in size to support a much larger number of pending accesses into the SSDs. %

%% file: challenges.tex
\section{Challenges}
\label{sec:challenge}
\pname{} proposes a paradigm shift in hardware and software stack. Of course this introduces several challenges that need to be addressed to make the \pname{} architecture a reality. 
In this section, we discuss some of the pressing challenges and provide some potential solutions.
 
 \subsection{Software Stack}
 One of the biggest challenges we have to tackle in \pname{} is the software stack.
 This is because traditional software is written to be run on the CPU, with the operating system providing routines, drivers and file systems to manage memory and storage devices and the data movement between them.
 An accelerator like the GPU is treated as an offload device in the traditional system, where the CPU moves the data to and from the accelerator's memory and launches compute kernels on the accelerator.
 However, in \pname{} we forgo many of these conveniences for higher performance.
 Thus, we have to rethink major components of the software stack.
 
 Although, we are able to leverage the \pname{} controller to handle some of the layers of the software stack, like NVMe drive management and the file system, many layers still remain a challenge.
 As an example, in the traditional system, the CPU uses the page fault mechanism in the operating system to lock pages and guarantee a consistent view of data in memory to many concurrent applications.
 However, managing a consistent and coherent page cache in a distributed environment like \pname{} without hurting performance is challenging.
 Thus, we leave it to the programmer to manage their own caches in the application's address space.
 Although this gives the application control over the use of its allocated memory, it restricts the sharing of such caches between applications.
 To provide sharing of cached data between applications would be a significant challenge for \pname{}.
 
 Furthermore, the programming model needed for a system like \pname{} is not clear.
 Traditionally, distributed systems leverage the divide and conquer approach and use message passing to communicate~\cite{mpi}. 
 However, for the emerging applications that require the view of the whole dataset, like certain graph processing workloads, message passing is not ideal.
 In our current design we use the CUDA parallel programming model for \pname{}.
 A shared memory system like CUDA requires memory abstractions with support for atomic operations to build complex communication primitives and leverages collective libraries like NCCL~\cite{nccl} to aggregate results.
 One of the big challenges in supporting such features in \pname{} is the fact that \pname{} enables the use of the NVMe protocol~\cite{nvme} from the application address space for low latency accesses.
 Thus, in \pname{} these communication primitives need to be built in the user-space, whereas in a traditional system the virtual memory, file system, and collectives libraries~\cite{nccl,mpi} can provide such abstractions over the underlying interconnect, memory and storage devices. We plan to provide a user-level library for the most common use cases.

To ease the programmer's burden with \pname{}, we can further remove the NVMe control from user-space, sacrificing performance, and use the NVMe SSDs as memory-mapped devices, with the \pname{} Controller providing the needed address translations.
However, non-volatile memories like SSDs introduce significant challenges for systems software due to the terabytes of capacity they offer.
Common system algorithms like memory reclamation and defragmentation, garbage collection, and page table walks
have been shown to break-down and not scale at large memory capacities~\cite{0sim}.
Although prior work has attempted to alleviate some of these system level issues by exploiting byte-addressable SSDs~\cite{2bssd, flatflash}, merging multiple indirection layers~\cite{flatflash}, and generally reducing software overheads~\cite{flashshare,FlashVM,spacejmp}, we are still a long ways from being able to provide a fully flat memory abstraction to the programmer in a system like \pname{} with low performance overheads.

 \subsection{Flash Memory}
Another challenge is in current SSD design.
Traditionally, to compensate the high Flash array access latency, NAND Flash greatly increased the data width so each access can potentially exploit spatial locality.
The Flash array data width of modern SSD is about 16KB~\cite{sandisk}.
However, for the real world workloads such as databases and graph analytics, 16KB is too large as the spatial locality is limited in these applications.
For example, Kavalanekar et al.~\cite{realstat} mention that the most common data request size in a real world database is 4KB.
In graph traversal applications, a few hundred bytes are needed per request~\cite{emogi}.
If we need only a fraction of data per Flash array access, moving the entire data out for every access in current SSD design wastes limited interconnect bandwidth.
To overcome such limitation, Samsung~\cite{issccznand} and Micron~\cite{snapread} proposed a partial Flash array access.
However such techniques still fetch large chunks of data, typically 8KB.
To truly maximize the random access bandwidth to the Flash array, it is necessary to enable a much smaller granularity of data access such as 128B.
With such fine-granular access and the interface provided by byte-addressable SSDs ~\cite{flatflash,2bssd}, Flash can more seamlessly be integrated in \pname{}'s memory system, without the overheads of the NVMe protocol.

\subsection{Interconnect}
Scalability in a system like \pname{} requires a high-throughput low-latency interconnect. 
Emerging interconnects such as NVLink~\cite{nvlink}, PCIe Gen5~\cite{pcisig}, CCIX~\cite{ccix} and Gen-Z~\cite{genz} provide an order of magnitude higher throughput over the current PCIe Gen3 interconnect. 
However, the performance of these interconnects drops when cacheline data packets are transferred over the network because of huge command overheads. 
For instance, PCIe Gen3 has a command overhead of $\sim$0.4$\%$ for 4KB transfer and $\sim$33.3$\%$ for cacheline-sized transfers.
The fundamental issue here is coming from header sizes being independent of data payload.
Changes in interconnect protocols are required to provide efficient cacheline level access to large memory with low overhead.
In addition to this, current interconnects support a limited number of outstanding requests. 
This is because these interconnects are designed assuming the CPU's low request-level parallelism. 
However with the \pname{} architecture, which can have millions of requests in flight, the interconnect protocols must be able to keep up. 
Furthermore, data placement and scheduling on \pname{}'s EPUs
need to be aware of and be optimized for the underlying interconnect.

%% file: conclusion.tex
\section{Conclusion}
\label{sec:conclusion}
In this paper, we highlight the fundamental challenges of modern computing systems for emerging applications such as machine learning, recommendation systems, graph and data analytics.
These applications require large memory capacities and high bandwidth for efficient execution.
However, even though architectural improvements have enabled the continued scaling of compute throughput, memory technology has failed to keep up.
Furthermore, we discuss the key limitations of the CPU-centric approach that is prevalent in system design.
This approach leaves the GPU, the modern computing device of choice, with very slow access to a limited pool of memory, contradicting the needs of emerging applications.
To address this, we rethink system design and propose the \pname{} system architecture.
\pname{} provides a scalable way to improve not only the compute capabilities of a system, but also the memory capacity and bandwidth, through near-data processing and highly-parallel Flash based memory systems.
Moreover, \pname{} removes the slow CPU and operating system from the access path of the compute units while still providing secure access to the non-volatile Flash memory.

%% file: acknowledgment.tex
\section*{Acknowledgement}
We would like to thank IMPACT group members, Isaac Gelado, Eiman Ebrahimi, and Steve Wallach for their invaluable technical advice and anonymous reviewers for their feedbacks.
This work was partially supported by the Applications Driving Architectures (ADA) Research Center and Center for Research on Intelligent Storage and Processing-in-memory (CRISP), JUMP Centers co-sponsored by SRC and DARPA, IBM-ILLINOIS Center for Cognitive Computing Systems Research (C3SR) - a research collaboration as part of the IBM AI Horizon Network. 
This work would not have been possible with the generous hardware donations from IBM, Xilinx and NVIDIA. Laslty,